# Giant shape-memory effect in twisted ferroic nanocomposites

Donghoon Kim[1]†, Minsoo Kim[1]†, Steffen Reidt[2], Hyeon Han[3], Hongsoo Choi[4], Josep Puigmartí-Luis[5,6], Morgan Trassin[7], Bradley J. Nelson[1], Xiang-Zhong Chen[1]*, Salvador Pané[1]*

[1] Multi-Scale Robotics Lab, Institute of Robotics and Intelligence Systems, ETH Zurich, Tannenstrasse 3, 8092 Zurich, Switzerland

[2] IBM Research Zurich, Säumerstrasse 4, 8803 Rüschilikon, Switzerland

[3] Max Plank Institute of Microstructure Physics, 06120 Halle (Saale), Germany

[4] Department of Robotics & Mechatronics Engineering, DGIST-ETH Microrobotics Research Center, Daegu Gyeong-buk Institute of Science and Technology (DGIST), Daegu, Republic of Korea

[5] Department of Physical Chemistry, University of Barcelona, Martí i Franquès, 1, 08028, Barcelona, Spain

[6] Institució Catalana de Recerca i Estudis Avançats (ICREA), Pg. Lluís Companys 23, Barcelona 08010, Spain

[7] Department of Materials, ETH Zurich, 8093 Zurich, Switzerland

* Corresponding authors. Email: chenxian@ethz.ch, vidalp@ethz.ch

† These authors contributed equally to this work

**Abstract:**

The shape recovery ability of shape-memory alloys vanishes below a critical size (~50nm), which prevents their practical applications at the nanoscale. In contrast, ferroic materials, even when scaled down to dimensions of a few nanometers, exhibit actuation strain through domain switching, though the generated strain is modest (~1%). Here, we develop free-standing twisted architectures of nanoscale ferroic oxides showing shape-memory effect with a giant recoverable strain (>10%). The twisted geometrical design amplifies the strain generated during ferroelectric domain switching, which cannot be achieved in bulk ceramics or substrate-bonded thin films. The twisted ferroic nanocomposites allow us to overcome the size limitations in traditional shape-memory alloys and opens new avenues in engineering large-stroke shape-memory materials for small-scale actuating devices such as nanorobots and artificial muscle fibrils.

**Main Text:**

The development of nanoscale machines, such as nanoelectromechanical systems, nanorobots, nanoscale aerial vehicles, and injectable miniaturized medical devices, places a pressing demand on nanoscale mechanical actuation architectures and materials. Shape-memory alloys (SMAs) have been widely explored as actuating materials because of their large deformation capability caused by reversible martensitic phase transformations. However, their application at the nanoscale is highly constrained because martensitic phase transformations are suppressed below a critical size (~ 50 nm) (*1, 2*).

Ferroelastic and ferroelectric oxides maintain their electromechanical response at the nanoscale, even in films of just a few atomic layers (*3*). Their electromechanical behavior is associated with changes in their crystalline lattice and ferroic domains. The switching of domains can induce recoverable strains and possibly shape-memory effect, whereas the mechanism is different from the martensitic phase transformation in SMAs (*4, 5*). At the nanoscale, the domains can be modulated by substrate-induced strain engineering to improve electromechanical response, yet macroscopic actuating strain barely exceeds 1% (*5-7*). Substrate removal could allow further tuning of the ferroic domains by changing boundary conditions and, thus, the exploration of unexpected material properties of the freestanding structures, such as superelasticity and large electromechanical response (*8-10*).

In this work, we demonstrate shape-memory effect with giant recoverable deformations (>10%) in freestanding architectures of ferroic oxide thin film by amplifying domain switching-induced strains through geometrical twist insertion. Twist insertion has been employed in polymeric coiled fibers and metamaterials to amplify strokes (*11-14*). However, it has never been explored in ferroic oxide ceramics, because of their brittle nature in bulk form, or because they are mechanically constrained to the substrate on which they are deposited. Notably, when the crystallite size scales down to the nanoscale, ceramic materials show high strength and large elastic strain endurance (*8, 9, 15-17*). This unique property, together with a bilayer design, enables us to realize predefined architectures that cannot be achieved in the single-layer freestanding approach (*3, 8*). The twisted architecture was fabricated by releasing patterned $BaTiO_3$ /$CoFe_2O_4$ (BTO/CFO) bilayer thin films from the substrate. Large shape-memory effect was observed through *in-situ* nanomechanical testing. The twisted architecture has a film thickness of ~ 20 nm, overcoming the size limitation encountered in conventional shape-memory alloys.

The twisted architecture was fabricated from BTO/CFO epitaxial bilayer thin films with a thickness of 8 nm and 15 nm, respectively (Fig. 1A). First, the BTO/CFO (001) bilayer was deposited onto a MgO (001) substrate using pulsed laser deposition (fig. S1). CFO was used to provide interfacial stress because of the large lattice mismatch (4 ~ 5%) between the two layers. The films were patterned into linear stripes of 1 μm wide and 70 μm long, with a tilt of 40° with respect to the [010] axis. After chemical etching of the MgO substrate (detailed information in the Methods section of the SI), the thin film stripes were released, and interfacial stress caused the film to roll around the [010] axis to form twisted architectures (Fig. 1B). Unlike brittle bulk ceramics, these freestanding twisted architectures exhibited a superelastic behavior. Although some structures were distorted by electrostatic and/or Van der Waals forces between the freestanding film and the substrate (fig. S2), they recovered their original shape after being mechanically detached from the substrate and exhibited spring-like behavior when pushed/pulled with a microneedle (Movie S1).

When the tensile stress is sufficiently large, the structure maintained the deformation, as can be seen from the changed helical pitch length in Fig. 1C. Interestingly, when the electron beam from the scanning electron microscope (SEM) was focused on the deformed structures, they recovered their initial shape (Fig. 1C, Movie S2), and their high superelasticity was maintained (fig. S3). This phenomenon is analogous to the conventional shape-memory effect, where deformed structures recover their original shape and mechanical properties via thermomechanical martensitic phase transformations (*18, 19*). However, the shape recovery in the twisted nanocomposite was triggered by the electron beam, i.e. electrical energy as opposed to thermal energy (*10, 20, 21*), which greatly facilitates the application of these structures at the nanoscale where localized thermal stimulation is not possible.

The shape-memory effect was also evaluated by *in-situ* nanomechanical tensile tests (Fig. 2A, fig. S4, and Movie S3). The structure was fully stretched with 9 μm elongation and a maximun tensile force of 1.5 μN. Upon complete unloading, the structure recovered. Figure 2B shows the non-linear force-displacement relationship of the twisted architectures obtained during the tensile loading and unloading processes. During the tensile test, there was energy dissipation ($E_{dissipation}$), which can be quantified by the area enclosed by the force-displacement curves. $E_{dissipation}$ increased gradually at a small strain level ($4.08 \times 10^{-3}$ J/cm$^3$ to $2.17 \times 10^{-2}$ J/cm$^3$ during 3.6 μm to 7.1 μm elongation), and increased abruptly when the twisted architecture approached the length limit ($2.17 \times 10^{-2}$ J/cm$^3$ to $1.92 \times 10^{-1}$ J/cm$^3$ during 7.1 μm to 9.1 μm elongation, Fig. 2C). The twisted architecture was tested more than one hundred cycles (9 μm stretching) with a gradual decrease in the enclosed areas ($E_{dissipation}$) of the force-displacement hysteresis loop (Fig. 2D, E). Interestingly, $E_{dissipation}$ 'recovered' upon electron beam irradiation (Fig. 2E). After 20 cycles of tensile tests, the structure was re-exposed to the electron beam for a fixed amount of time, and the tensile test was resumed after turning off the electron beam. Notably, $E_{dissipation}$ increased ~ 0.05 J/cm$^3$ after the beam exposure, and this 'recovery' of the dissipated energy was observed after every re-exposure. When we compared $E_{dissipation}$ during tensile test cycles as a function of the beam current under the continuous beam irradiation condition (fig. S5), we observed that more energy was consumed with a higher beam current. Additionally, the structure maintained the 'recovered states' (i.e. high-energy-consuming state) under continuous beam irradiation (after 4th cycle in Fig. 2E).

From the phenomenological Landau-Ginzburg theory, the elastic strain in ferroelectric materials is expressed as a function of the ferroelectric polarizations:

$$e_{ij} = \varepsilon_{ij} - Q_{ijkl}P_kP_l \qquad (1)$$

where $\varepsilon_{ij}$ is the total local strain and $Q_{ijkl}$ is the electrostrictive coefficient (*22*). Therefore, the change of ferroelectric polarization alters the strain in ferroelectric materials and vice versa. Meanwhile, the electron beam irradiation on ferroelectric materials leads to the accumulation of surface charges, which can eventually switch or rotate ferroelectric polarizations to the out-of-plane direction (*23-26*). Here, we define the domains with polarization direction in the film plane (or along the tangent of the film plane when the film is curled) as in-plane domains, and those with the polarization direction perpendicular to the tangent of the film plane as out-of-plane domains. As the complex crystalline orientation variations within the twisted architecture make the experimental access to the domain configuration unfeasible, we performed phase-field model simulations (detailed information in the Methods section of the SI) based on the time-dependent Landau-Ginzburg equation (*27, 28*). Figure 3A depicts the BTO ferroelectric domains in freestanding BTO slabs. In the initial state, only the bottom surface was tensile-strained, representing strain coming from the CFO layer, with a domain configuration similar to that of the bent freestanding BTO films (*8, 29*). When tensile stresses were applied on both sides of the slab (loading tensile force), the slab was mechanically stretched and the in-plane domain density increased. As negative charges accumulated on the slab surface during the electron beam exposure, ferroelectric polarization switched towards an out-of-plane orientation, giving rise to the change of strain profile and the retraction of the mechanically stretched BTO slab (shape recovery, fig. S6). Therefore, it is likely that the switching between the in-plane and out-of-plane oriented ferroelectric domains, induced by mechanical stress and the electric field, is responsible for the observed shape deformation and its recovery.

Based on our findings, we suggest a possible mechanism for superelasticity and the shape-memory effect in the twisted architectures (Fig. 3B). As an epitaxially strained ferroelectric material, BTO possesses a complex domain configuration, where domain switching enables recoverable strains (*4, 8, 30, 31*). Shape deformations and recoveries were achieved by the interplay between the stress induced by the ferroelectric domain switching in the BTO layer and the mechanical stress imposed by the bottom CFO layer. In the low-strain regime, the twist shows a superelastic response (①→②→① in Fig. 3B). While surface tension-modulated elastic deformation cannot be ruled out (*32*), ferroelectric polarization switching undoubtedly contributes to the superelastic response. The ferroelectric polarization switches to an in-plane direction during stretching and returns to the original direction during release, as shown by the simulation results and as observed elsewhere (*8*). The deformation is maintained in the large strain regime and even beyond the threshold strain (②→③→④ in Fig. 3B). This can be attributed to either the residual stress or domain pinning during mechanical loading (*33-35*). When the electron beam was focused on the structure, the ferroelectric polarization switched from in-plane to out-of-plane, providing the force for shape recovery (④→⑤→① in Fig. 3B). The in-plane domain pinning can be corroborated by the gradual decrease in $E_{dissipation}$ with an increased cycling number (Fig. 2E), although the movement and/or accumulation of the defects, such as dislocations in the structure, could also contribute to the decrease of $E_{dissipation}$ (*36, 37*). However, as the $E_{dissipation}$ recovered after the beam exposure and the $E_{dissipation}$ increased when the electron beam was focused on the architecture, this clearly demonstrates the involvement of the polarization-switching modulated mechanism. We tested the repeatability of this shape-memory response and irrespective of the type of external force that was used for deformations, the shape-memory effect was consistently observed with electron beam irradiation (movies S4, S5).

Compared to conventional SMAs where martensitic phase transformation is suppressed below the critical size, a domain switching-induced shape-memory effect in ferroic oxides was observed in films of ~20 nm (Fig. 4A). This is the smallest feature size at which shape-memory effect has been demonstrated. Our results can also be extrapolated to smaller scales, as ferroelectricity occurs even at the monolayer (*3*). In addition, the shape recovery of twisted architecture is driven by electric fields with remarkable recoverable strains (Fig. 4B). The twisted architectures may have the potential to be actuated with other external stimuli, such as temperature and magnetic fields, since (i) BTO goes through phase transformations at Curie temperatures, and (ii) CFO is ferromagnetic.

In summary, we designed and demonstrated ferroelectric domain switching-induced shape-memory twisted architectures by changing the boundary condition via releasing the film from the substrate and introducing geometric engineering. An electrically induced large recoverable strain was achieved in the twisted architectures. Shape memory effects realized through this approach can bypass the critical size limitation encountered in the conventional SMAs. We believe that our discovery will enable new large-stroke shape-memory materials and structures for small-scale devices, such as micro- and nanorobots, actuators, and artificial muscles.

**Acknowledgement**

This work has been financed by the ERC Consolidator Grant "Highly Integrated Nanoscale Robots for Targeted Delivery to the Central Nervous System" HINBOTS under the grant no. 771565, the MSCA-ITN training program "mCBEEs" under the grant no. 764977, the ERC Advanced Grant "Soft Micro Robotics" SOMBOT under the grant no. 743217, and the Swiss National Science Foundation (Project No. 200021L_192012). X. C. would like to acknowledge the Swiss National Science Foundation (No. CRSK-2_190451) for partial financial support. This work was partially supported by the National Research Foundation of Korea (NRF) (No. 2021M3F7A1082275 and No. 2017K1A1A2013237), funded by the Ministry of Science and ICT of Korea. M. T. acknowledges the financial support by the Swiss National Science Foundation under project No. 200021_188414. M. K. acknowledges partial financial support by the Swiss National Science Foundation under project No. 200021L_197017.The authors would also like to thank the Scientific Center for Optical and Electron Microscopy (ScopeM) and the FIRST laboratory at ETH for their technical support, and the Cleanroom Operations Team of the Binning and Rohrer Nanotechnology Center (BRNC) for their help and support.

**Figures**

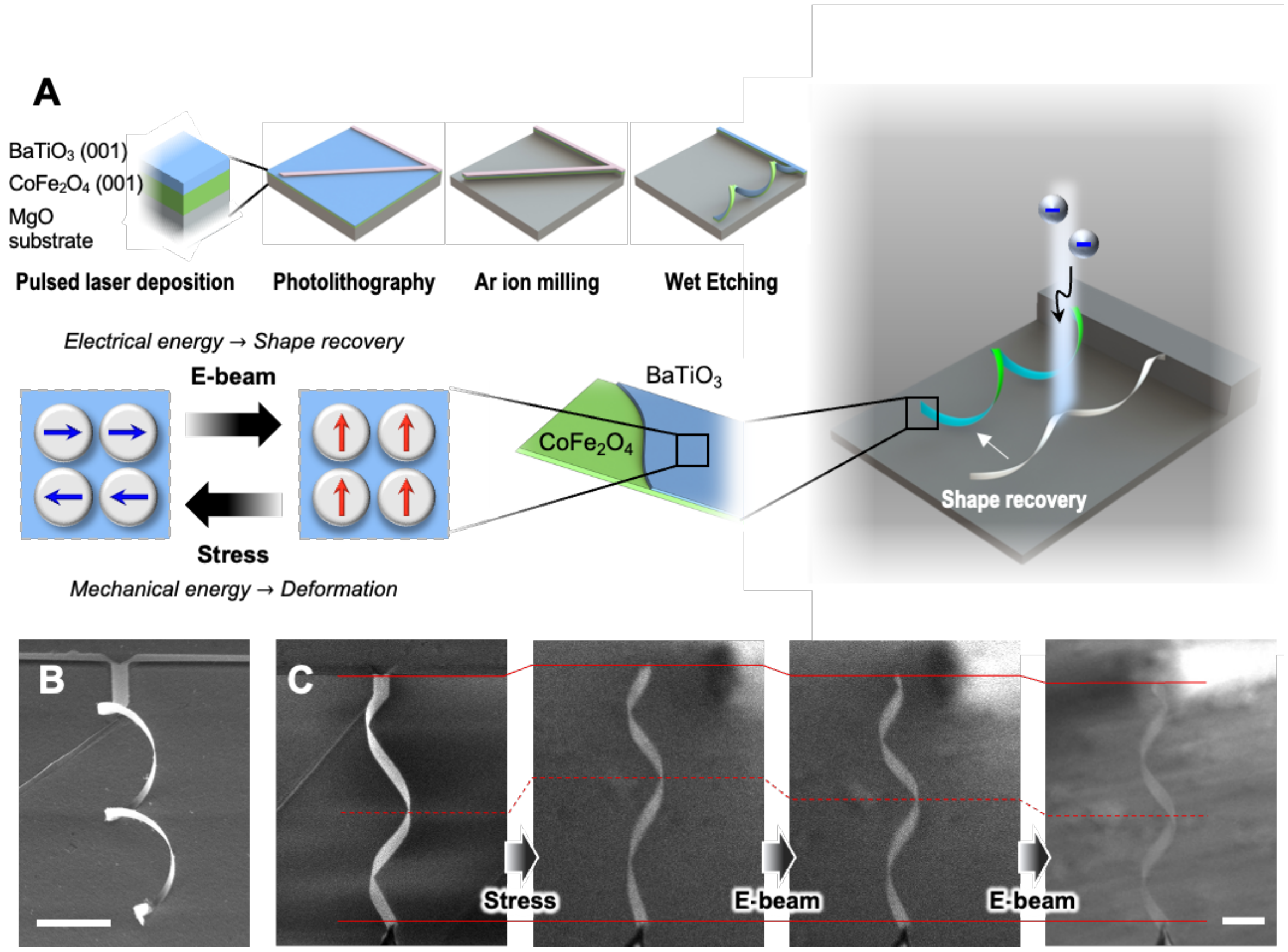

**Fig. 1. Electron beam induced shape-memory effect in twisted BTO/CFO nanocomposites.** (**A**) Schematic diagram of the twisted BTO/CFO fabrication process and the shape-memory effect under electron beam irradiation. (**B**) SEM image (45° tilted) of the fabricated twisted BTO/CFO nanocomposite. (**C**) Deformation under tensile loading and shape-memory behavior under electron beam irradiation. Scale bars indicate 10 µm.

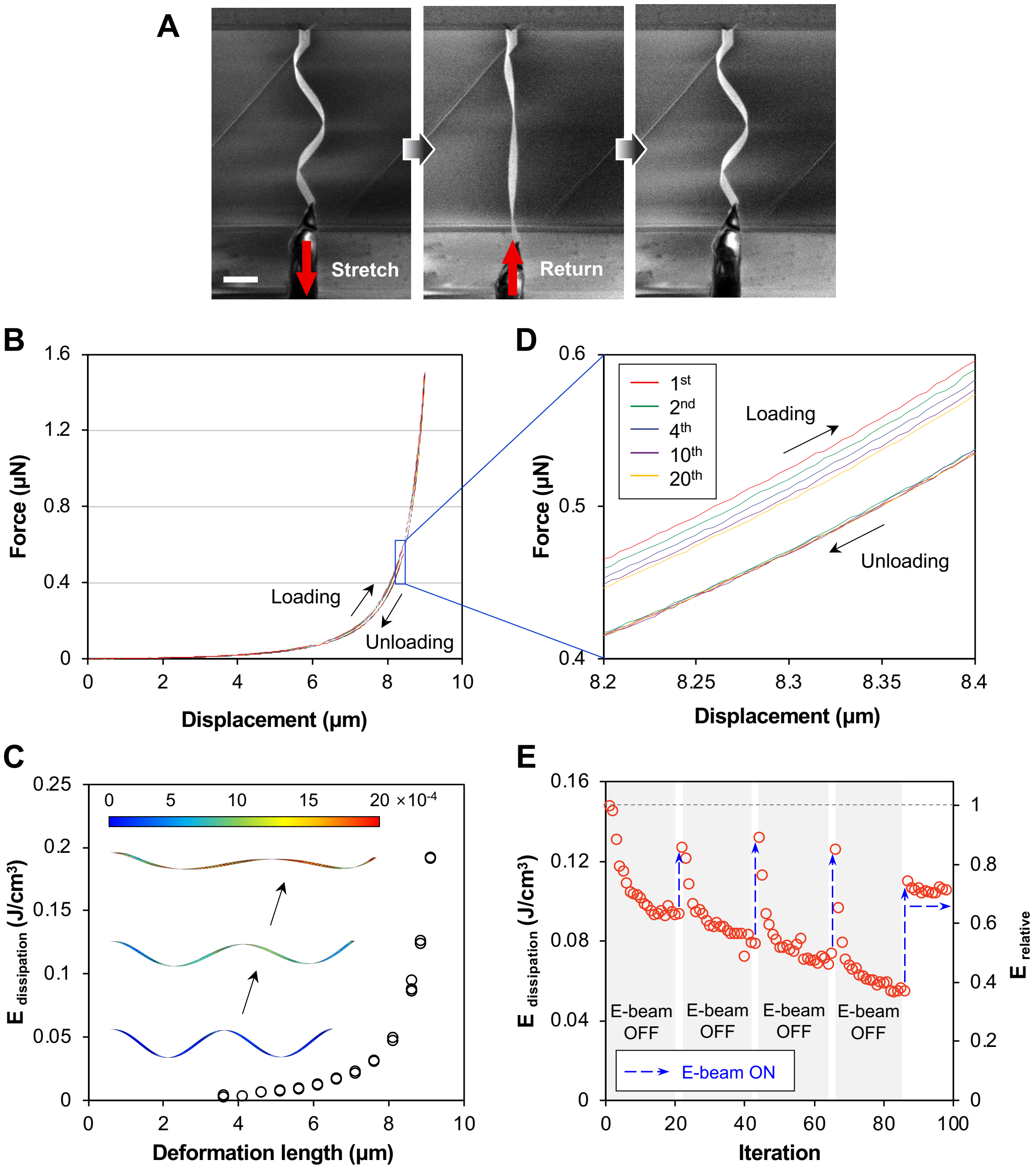

**Fig. 2. *In-situ* nanomechanical testing.** (**A**) Sequential SEM images of the tensile test of the twisted BTO/CFO (scale bar: 10 μm). (**B**) Non-linear force-displacement curve measured during tensile loading and unloading. (**C**) $E_{dissipation}$ as a function of the deformation (stretching) length. Inset images show the strain evolution during stretching, estimated by a finite-element method structural analysis. $E_{dissipation}$ increases exponentially as the twisted architecture is more strained. (**D**) Magnified force-displacement curves showing the degradation over the repetitions. (**E**) $E_{dissipation}$ change over the tensile cycling tests. $E_{dissipation}$ degraded over iteration and recovered under electron beam irradiation. Recovered state was maintained under continuous irradiation.

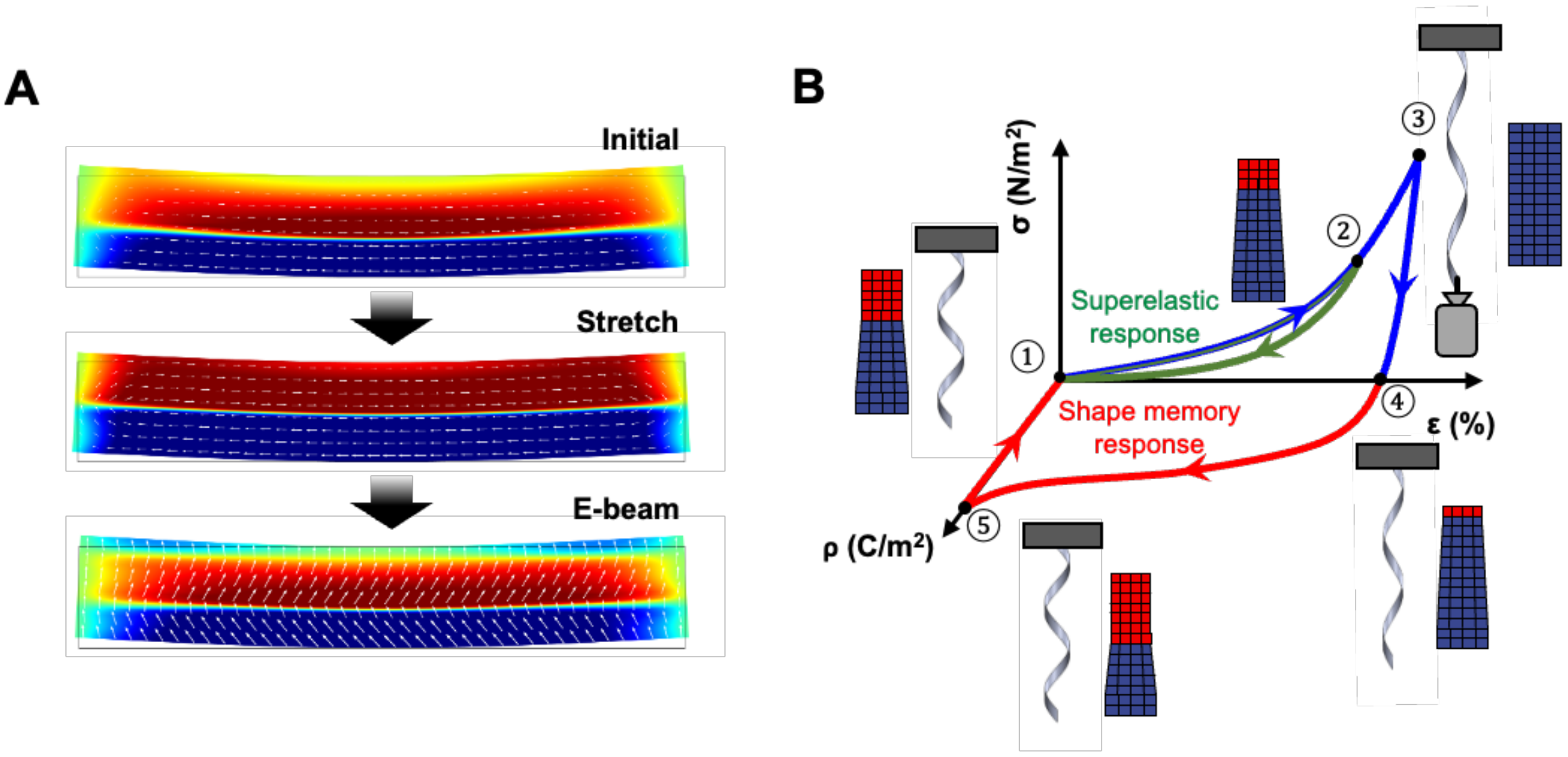

**Fig. 3. Phase field model simulation on the effect of mechanical stress and electron beam.** (**A**) In-plane ferroelectric domain mapping in the BTO slab. As tensile stress was applied in the BTO slab, the in-plane domain density increased. The ferroelectric polarizations then switched to out-of-plane directions under surface electrical charging. White arrows indicate the polarization directions. (**B**) Proposed cyclic behavior of the superelastic and shape-memory responses with schematic representation of the corresponding ferroelectric domain switching. The blue grids represent in-plane domains, and the red grids represent out-of-plane domains.

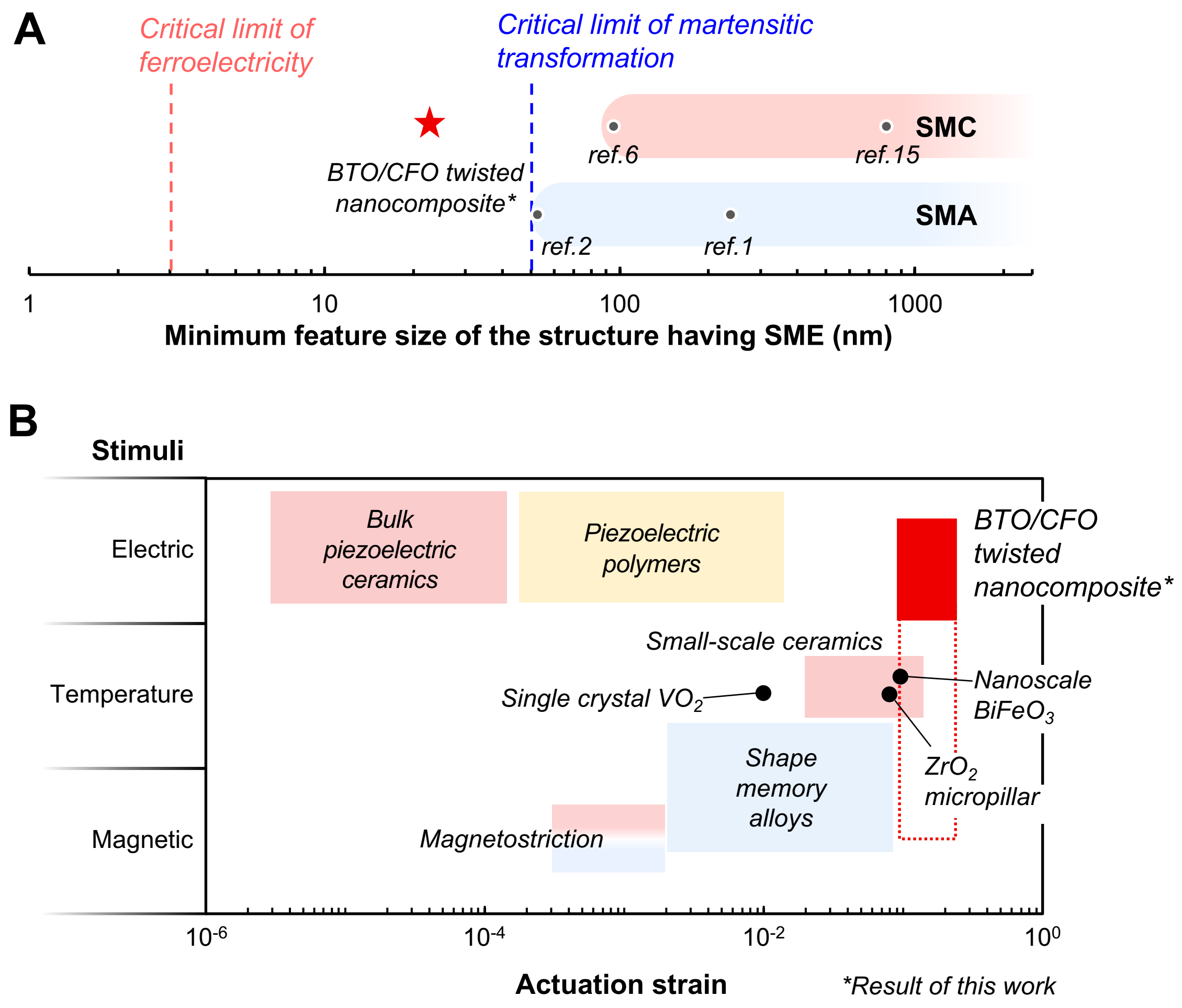

**Fig. 4. Comparison of feature size and actuation strain in different materials.** (**A**) The minimum feature size of materials with shape-memory effect (SME). For martensitic phase transformation-based shape-memory alloys (SMAs) and ceramics (SMCs), the critical feature size is limited to apporoximately 50 nm as the phase transformation is suppressed below this limit. For shape-memory ferroelectrics, however, ferroelectric domain switching can occur down to a few nanometers, extending the minimum feature size to a smaller scale. (**B**) Comparison of the actuation strain in different materials grouped by the actuation stimulus (Red: ceramics, blue: metals, yellow: polymers). In a BTO/CFO twisted nanocomposite, the shape-recovery actuation can achieve more than 10%. Although the recoverable strain from the ferroelectric domain switching in the BTO is around 1%, the shape-memory effect is amplified by the structural design, giving a large actuation strain range.

# Supplementary Materials

## Methods

### Thin film deposition

BTO/CFO epitaxial thin films were grown on (001) oriented MgO single crystalline substrate (Crystal GmbH) using pulsed laser deposition with a 248 nm KrF excimer laser, as previously reported (*1*). MgO substrate was cleaned with acetone and ethanol under ultrasonic wave for 5 min, respectively, before the deposition process. A CFO layer was first deposited at 550 ℃, 10 mTorr oxygen partial pressure, and laser parameters of 5 Hz, 1.8 $J/cm^2$. Subsequently, a BTO layer was deposited at 750 ℃, 200 mTorr oxygen partial pressure with 4 Hz, 1.2 $J/cm^2$ laser. Theta-2theta X-ray diffraction and reciprocal space mappings show the epitaxial nature of as-deposited thin films with relaxed strain status of the BTO layer (fig. S1).

### Twisted BTO/CFO fabrication

Photoresist (AZ 1505) was spin-coated onto the BTO/CFO thin films, and arrays of tilted lines with 1 μm width, 70 μm length, and 40° angle from [010] axis were patterned using UV-photolithography on the film. Subsequently, the patterned film was dry-etched with Ar-ion milling (Oxford IonFab 300 Plus) with a 500–mA beam current for 20 s, for a total 15 times with 90 s resting between each milling session to prevent the photoresist from burning. After the dry-etching, the remaining photoresist was rinsed-off with acetone (5 min), IPA (5 min), and oxygen plasma (600 W, 3 min). Afterwards, the MgO substrate was wet-etched with sodiumbicarbonate saturated solution (*2*) and the sample was dried with a critical point dryer (Tousimis-CPD).

### *In-situ* nanomechanical tensile testing

Tensile tests were performed under a scanning electron microscope (Nova NanoSEM 450, FEI Company) equipped with a nanomechanical testing system (FT-NMT03, Femtotools AG) at room temperature (fig. S5). Twisted BTO/CFO nanocomposites were attached to the force sensor probe using SEM-compatible glue (SEMGLU, Kleindiek Nanotechnik GmbH), and the force was measured with 100 Hz sampling frequency under a 0.5 μm/s tensile loading-unloading rate using a micro-electro-mechanical system (MEMS)-based force sensor (model FT-S200). The force sensor has a tungsten probe tip with a radius of less than 0.1 μm, ± 200 μN force range limit, and 5 nN resolution.

Phase field Modelling

Ferroelectric polarization switching behavior was investigated using phase-field modelling. For simplicity, two-dimensional BTO slabs (60 nm × 10 nm) with different mechanical and charge boundary conditions were simulated by solving the time-dependent Ginzburg-Landau equation,

$$\frac{\partial P_i(r,t)}{\partial t} = -L\frac{\delta F}{\delta P_i(r,t)}, i = 1,2,3 \qquad (1)$$

where $P_i(r,t)$ is the polarization at location $r$ and time $t$, $L$ is a domain wall mobility related kinetic coefficient, and $F$ is the total free energy that can be expressed as

$$F = \iiint \left(f_{bulk} + f_{elec} + f_{grad} + f_{elas}\right) dV \qquad (2)$$

and the Landau free-energy density ($f_{bulk}$), electric energy density ($f_{elec}$), gradient energy density ($f_{grad}$), and elastic energy density ($f_{elas}$), are described by

$$f_{bulk} = \alpha_{ij}P_iP_j + \alpha_{ijkj}P_iP_jP_kP_l + \alpha_{ijkjmm}P_iP_jP_kP_lP_mP_n \qquad (3)$$

$$f_{elec} = -\frac{1}{2}\varepsilon_0\varepsilon_b E_iE_j - E_iP_i \qquad (4)$$

$$f_{grad} = \frac{1}{2}G_{ijkl}P_{i,j}P_{k,l} \qquad (5)$$

$$f_{elas} = \frac{1}{2}C_{ijkl}\left(\varepsilon_{ij} - \varepsilon_{ij}^{o}\right)\left(\varepsilon_{kl} - \varepsilon_{kl}^{o}\right) \qquad (6)$$

where $\alpha$'s are the Landau expansion coefficients, $\varepsilon_0$ and $\varepsilon_b$ are the vacuum permittivity and dielectric constant, $E_i$ is the electric field including both external field and depolarization field, $G_{ijkl}$ is the gradient energy coefficient, $P_{i,j}$ is the polarization gradient, $C_{ijkl}$ is the elastic stiffness tensor, and $\varepsilon_{ij}$ and $\varepsilon_{ij}^{o}$ are the total and spontaneous strain, respectively. The coefficients for the equations were adopted from the previous research (*3, 4*). Eq. (1) was transformed into general partial differential equation forms in finite-element method software COMSOL Multiphysics and numerically solved with 1 nm × 1 nm mesh size. Surface tension (which represents the interfacial strain from the CFO at the bottom surface of the BTO slab) and boundary loads (tensile forces on both sides of the slab) were considered as mechanical boundary conditions and a closed loop electrical boundary condition was adopted.

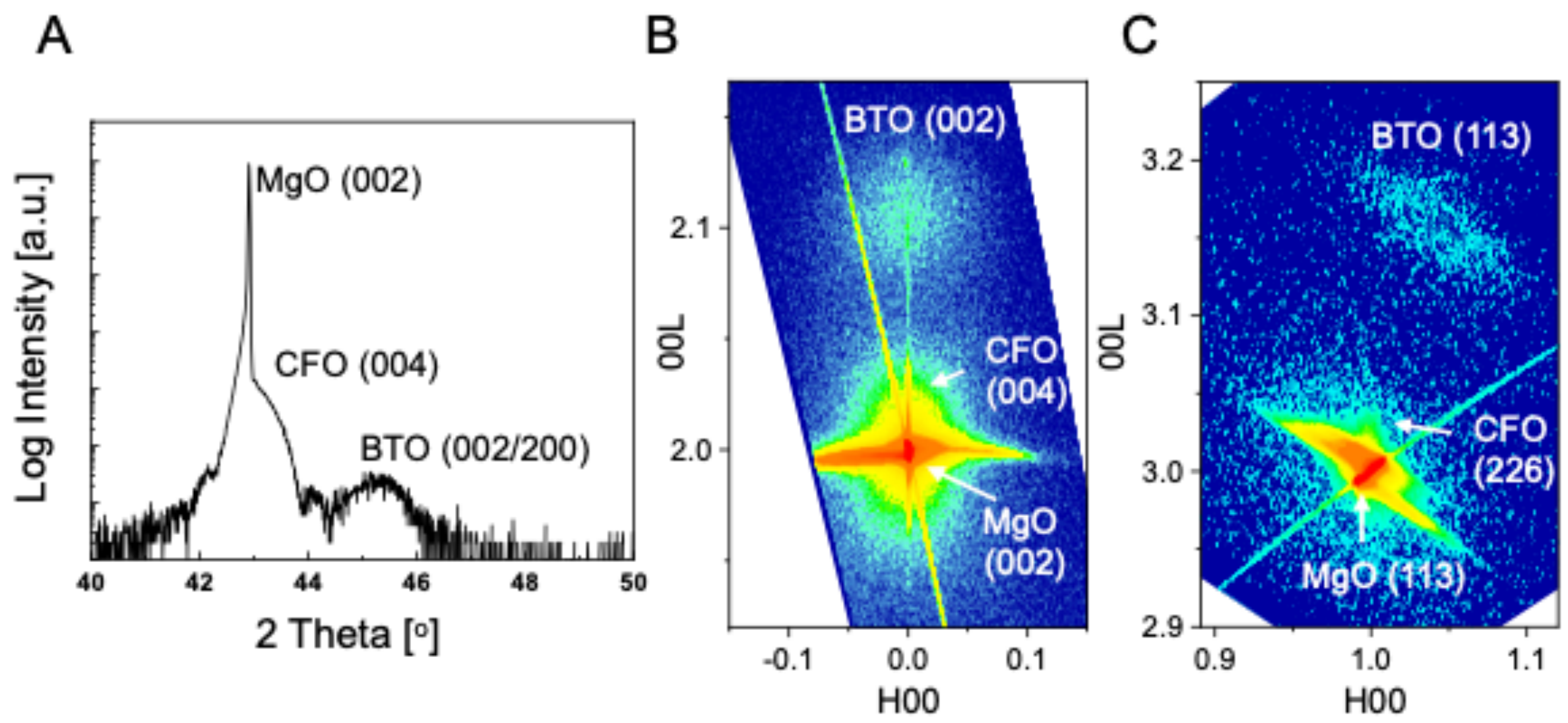

**Fig. S1. Epitaxial growth of BTO/CFO bilayer thin films.** (A) Theta-2theta scan of the BTO/CFO epitaxial thin film grown on MgO (001) substrate. (B-C) Reciprocal Space Mappings of the BTO/CFO//MgO thin film around MgO (002) and (113) Bragg peaks show epitaxial characteristics of the bilayer thin film with the relaxed strain state of the BTO layer.

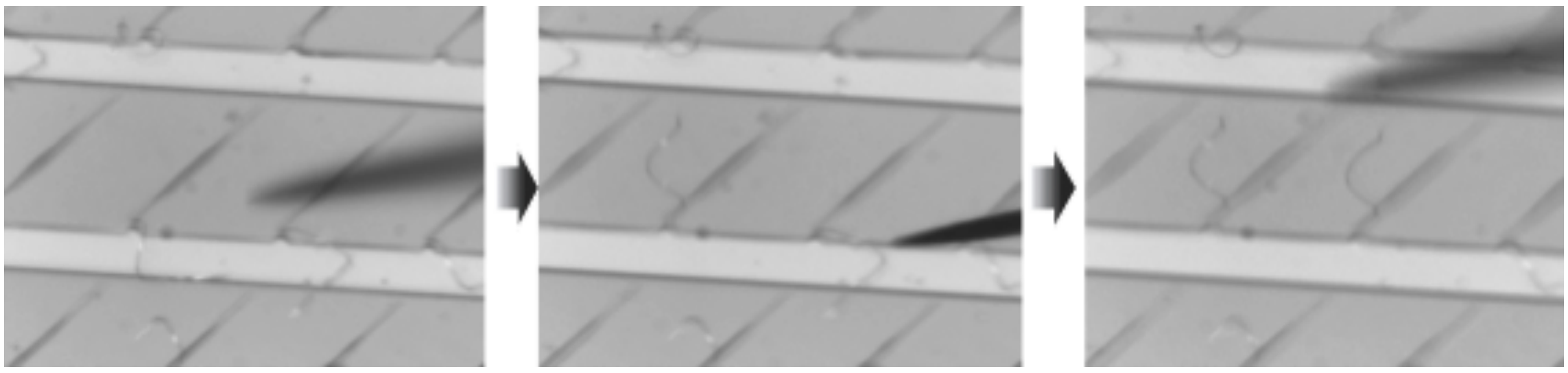

**Fig. S2. Superelasticity of twisted BTO/CFO nanocomposites.** Full shape recovery of the distorted structures into twisted architectures when detached from the substrate. A full video of the detachment and the pulling and stretching process is available in Movie S1.

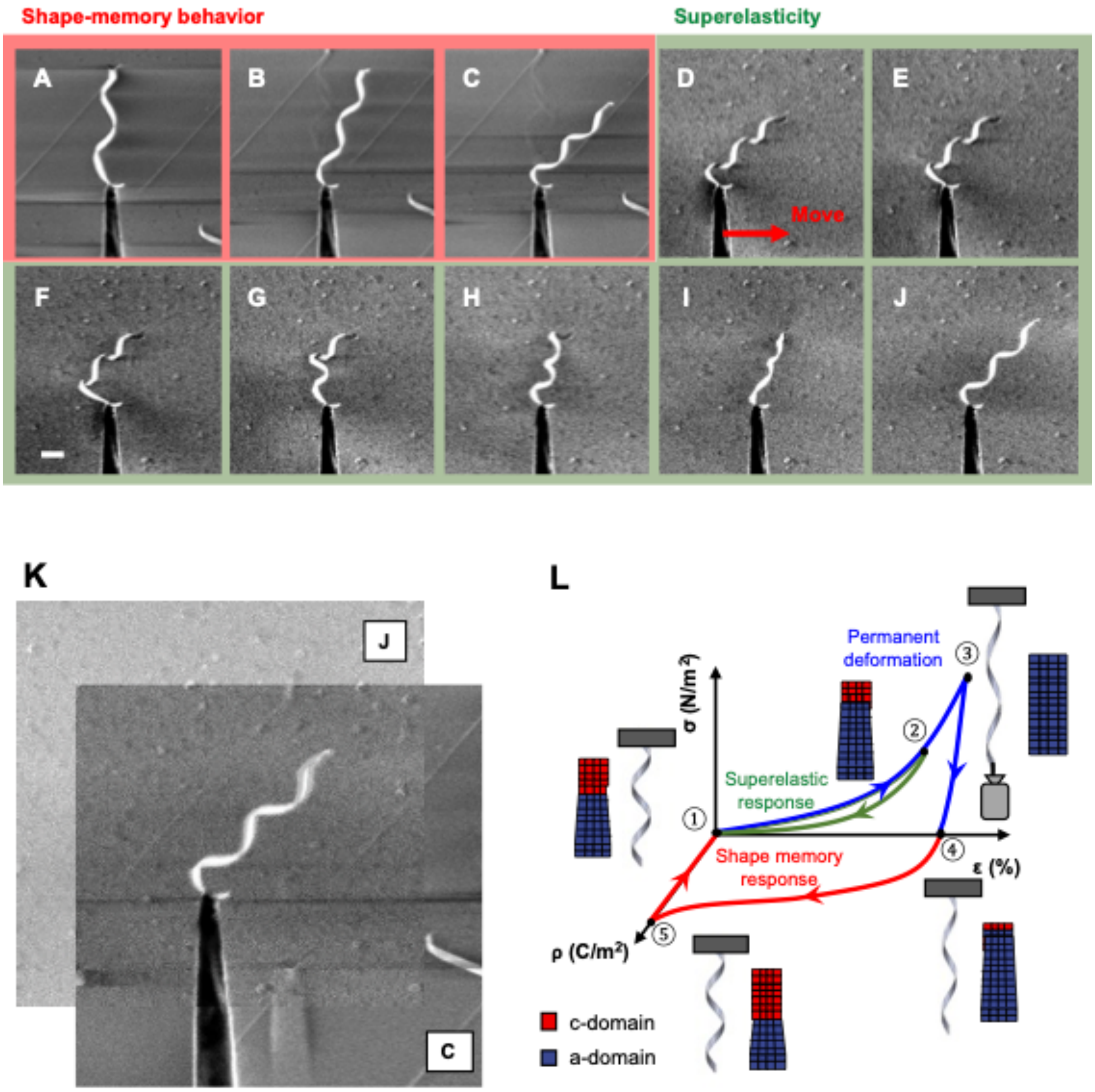

**Fig. S3. Maintaining superelasticity after showing shape-memory response under electron beam irradiation.** (A) The structure was stretched with the force sensor probe (1→2). (B) When enough tensile load was applied, the twisted BTO/CFO was fractured and the deformation was maintained (3→4). (C) Shape-memory effect when irradiated with electron beam (4→5→1). (D-I) Superelasticity of the recovered structure was tested using Van der Waals force between the substrate and the twisted BTO/CFO. Superelasticity was preserved after the shape memory behavior. (J) The shape was recovered after the electron beam irradiation. (K) Comparison of the physical shapes between (C) and (J) (before and after the superelasticity test), indicating full recovery of the twisted architecture. (L) Superelastic and shape-memory response cycle in twisted BTO/CFO with corresponding domain switching in the BTO layer.

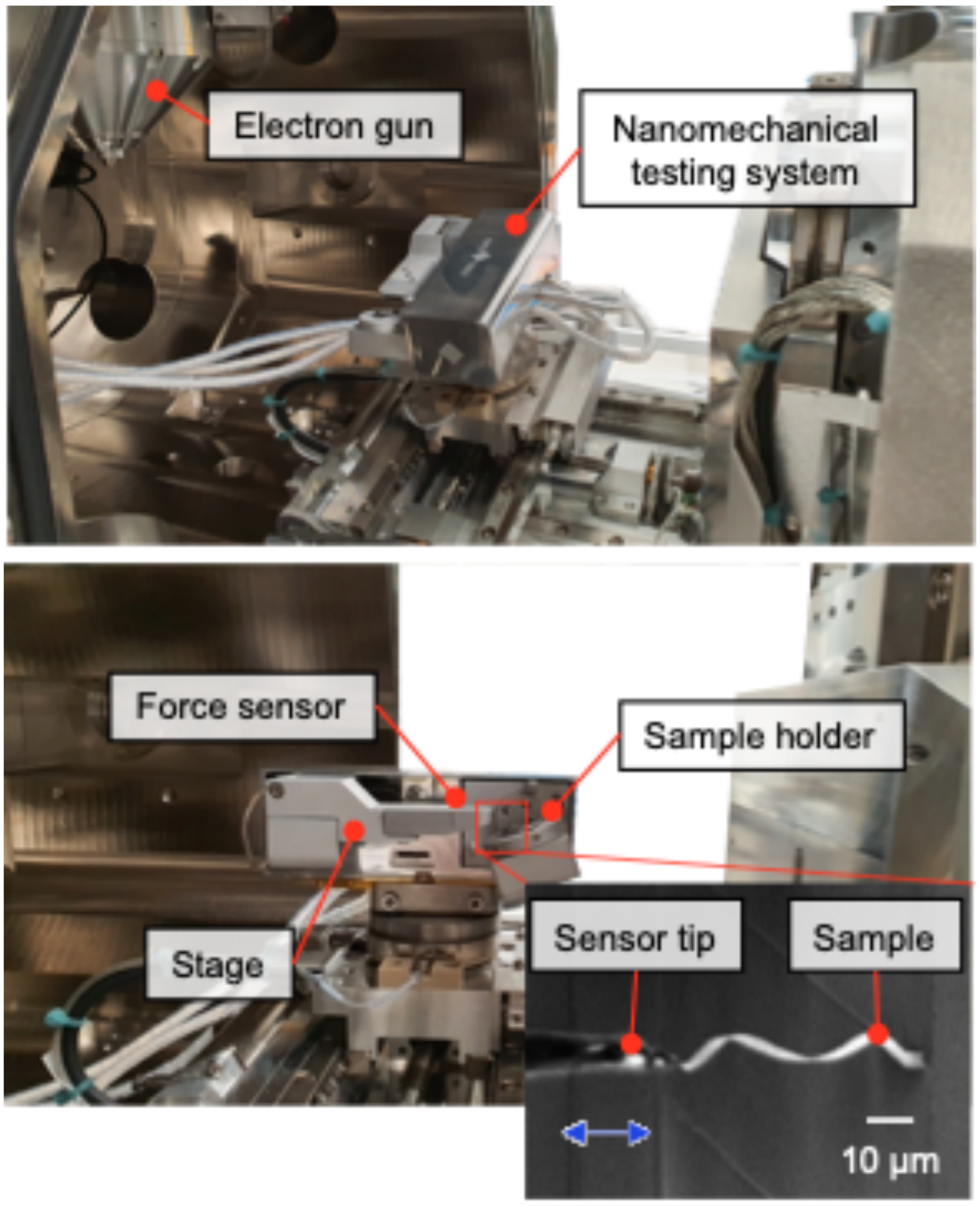

**Fig. S4.** ***In-situ*** **nanomechanical measurement setup (Femtotools) in SEM.**

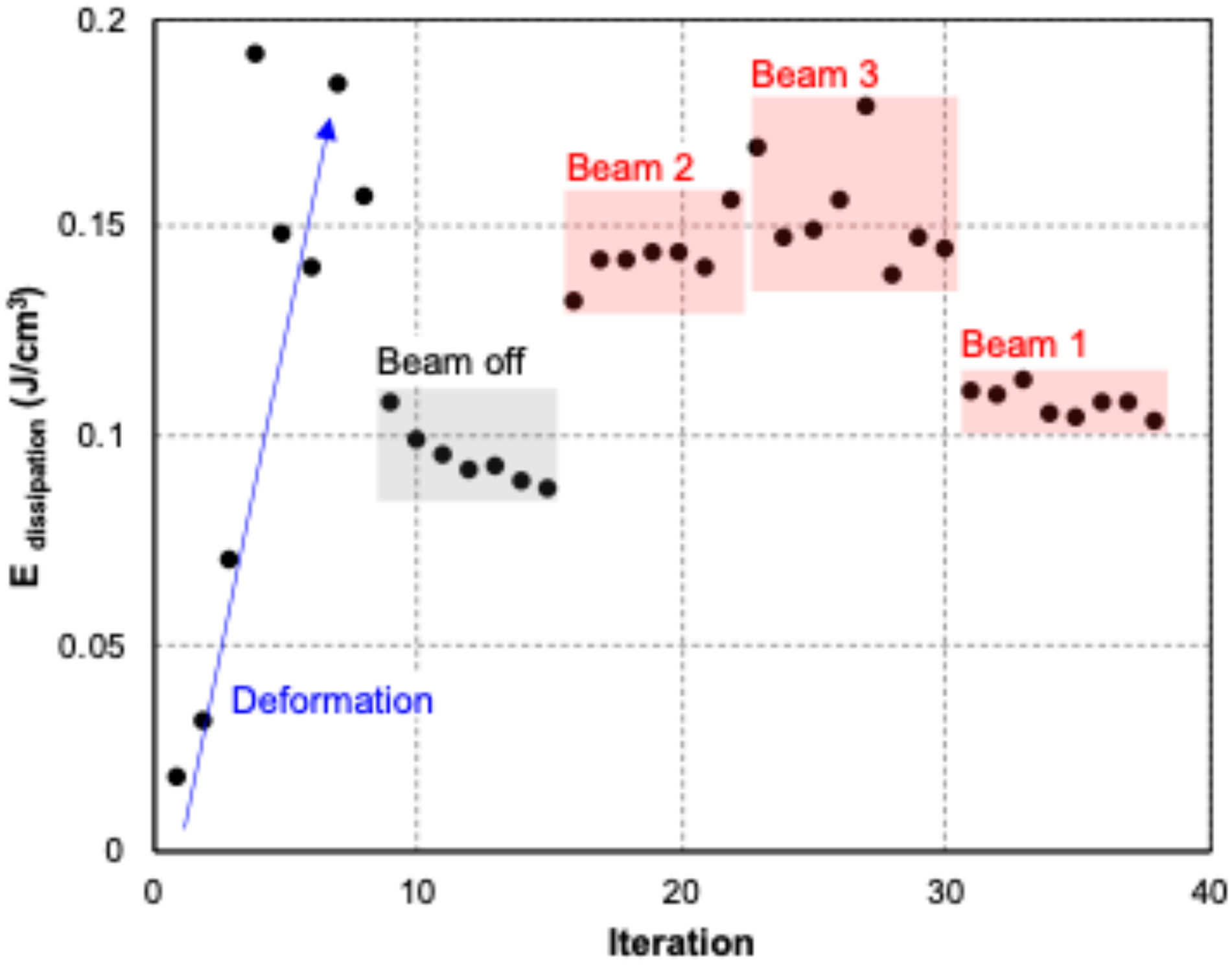

**Fig. S5. Electron beam current (dose) dependence of the $E_{dissipation}$.** Beam 1, 2, 3, indicates 30 pA, 100 pA, and 400 pA beam current, respectively, while maintaining the same magnification. With a stronger electron beam current, higher energy dissipation was obtained.

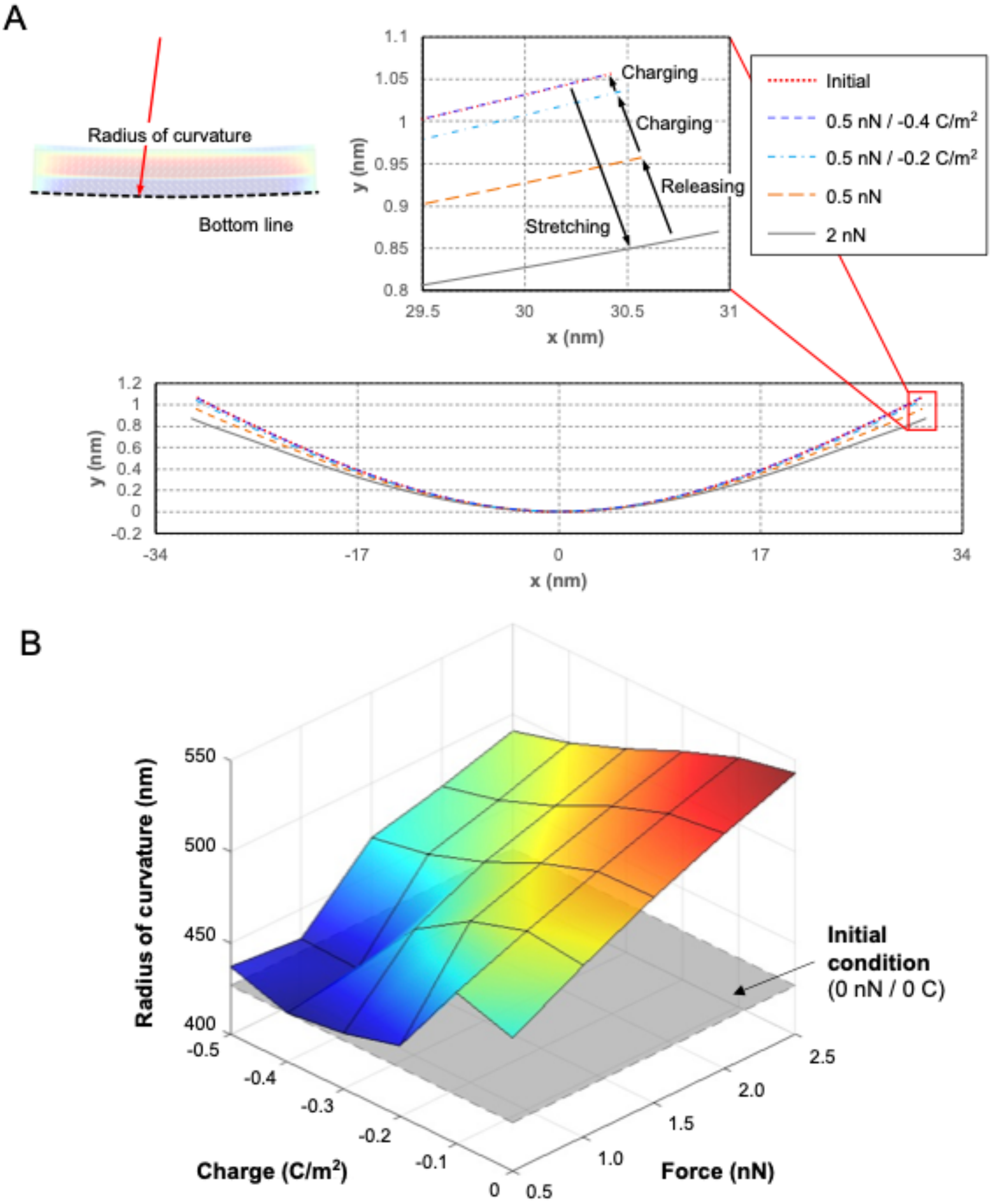

**Fig. S6. The effect of the tensile force and surface charge accumulation on physical shape of the BTO slab calculated with phase-field modelling.** (A) Displacements of the bottom surface of the BTO slab are plotted under different tensile force and surface charge boundary conditions. The values of the tensile forces and surface charges were assumed to be in states of stretching, releasing, and charging. Higher tensile forces resulted in more stretching in the x-direction, while surface charge accumulation resulted in a physical shape recovery to the initial

state. Here, the initial state refers to the bent BTO slab, where the bottom surface was tensile-strained due to the lattice mismatch between the BTO and the CFO. (B) Calculated radius of curvature of the bottom surface as a function of surface charges and stretching forces. As more surface charges are accumulated, the radius of curvature value gets closer to the initial condition, indicating shape recovery to the original state.